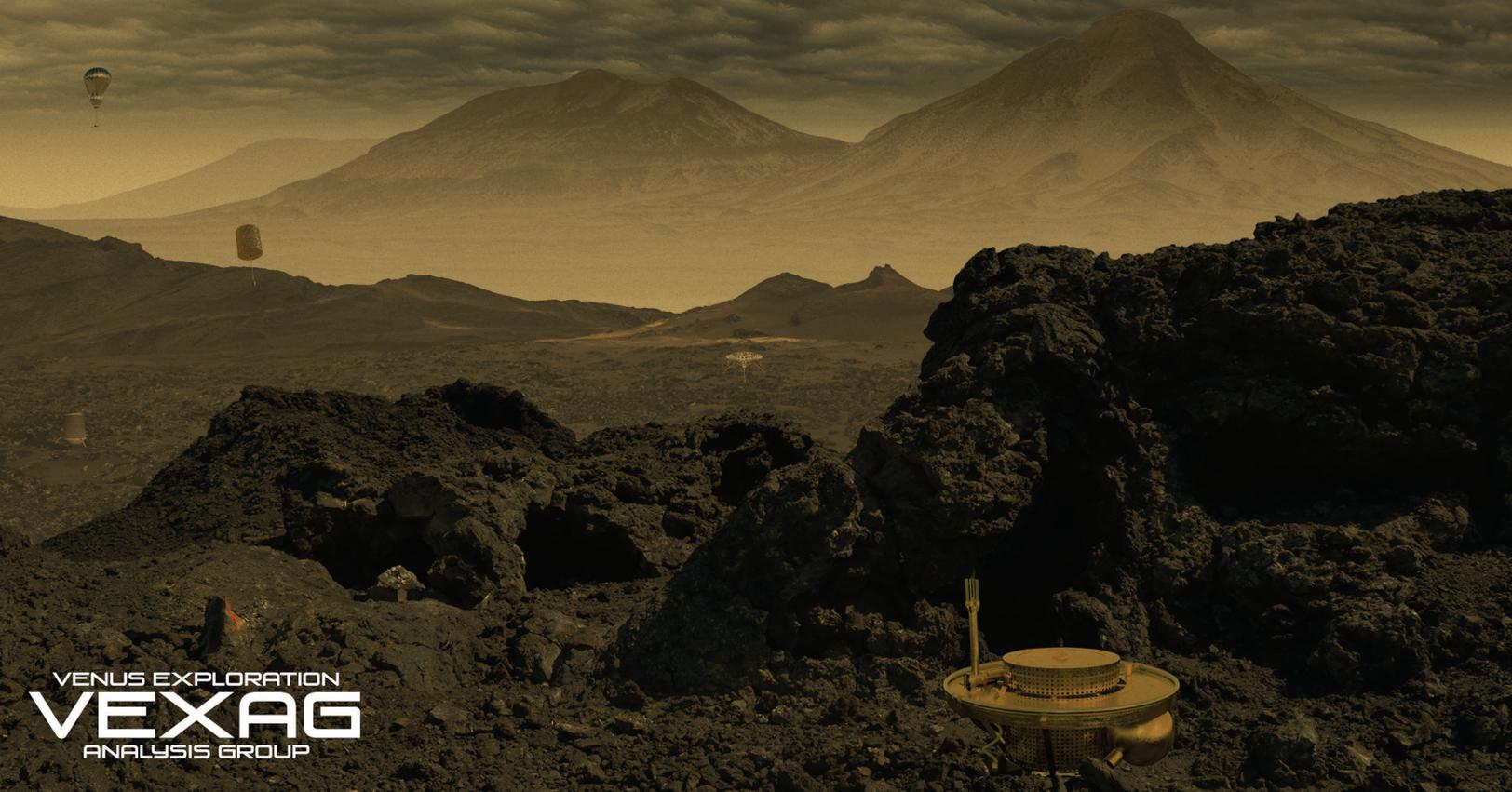

# A NEW STRATEGY FOR THE EXPLORATION OF VENUS

VENUS EXPLORATION
**VEXAG**
ANALYSIS GROUP

# EXECUTIVE SUMMARY

In early 2023, the Venus Exploration Analysis Group began development of a **new Venus exploration strategy** intentionally broader in scope, and looking to a longer time horizon, than standard VEXAG publications.

This strategy comprises **six complementary, related themes** for advancing Venus exploration:

Implementing this strategy will **lay the foundation for NASA establishing in the 2030s a formal Venus Exploration Program with a clearly defined end goal**.

There is a **compelling case** for a formal Venus Exploration Program:

- ✓ Venus is a **unique exploration target** where Decadal-level questions of interest to a wide range of planetary scientists can be tackled
- ✓ Venus is accessible, with **frequent launch windows and short cruise times**
- ✓ Its proximity means a **rapid science return** at a relatively low price point
- ✓ The next generation of Venus in situ platforms are **already technically mature**
- ✓ A series of planned, directed missions will **maximize the scientific value of Venus** and allow each new mission to synergistically build upon the last

Key to implementing this exploration strategy is conveying the scientific importance, value, and need for a clearly defined VEP to as broad a set of stakeholders as possible between now and the start of the next planetary science Decadal Survey.

**The work of establishing a dedicated Venus Exploration Program should start now.**

# PREFACE

[1] The Origins, Worlds, and Life Decadal Survey was published in spring 2022. It is the third such community consensus study report organized by the National Academies of Sciences, Engineering, and Medicine for NASA.

The 2023–2032 Planetary Science and Astrobiology Decadal Survey *Origins, Worlds, and Life*[1] recommended that "NASA should develop scientific exploration strategies, as it has for Mars, in areas of broad scientific importance, e.g., Venus…, that have an increasing number of U.S. missions and international collaboration opportunities" (*OWL*, p. 22-10).

In NASA's initial responses to that Decadal Survey, the agency asserted that "…specific scientific exploration strategies should be community generated by bodies such as the Analysis Groups," **thus placing the onus on the planetary community to generate and support these exploration strategies**.

In late 2022, the Venus Exploration Analysis Group thus began a project to develop a new exploration strategy for Venus, reflecting both the 2021 selections of the VERITAS, DAVINCI, and EnVision missions and the sweeping comparative planetology recommendations relevant to Venus in *Origins, Worlds, and Life*.

**This is that strategy.**

Taking a broad look at the scientific, technological, and programmatic advances required to address the key outstanding questions that Venus poses, **and predicated on VERITAS, DAVINCI, and EnVision flying as planned in the early 2030s**, this report outlines a set of actions available to NASA, VEXAG, and the planetary science community at large to establish a sustained program of Venus exploration in the years and decades ahead. Key to this approach is recognizing Venus as a unique setting where multiple, cross-disciplinary, Decadal-level planetary, Earth, heliophysics, and exoplanet science questions can be addressed—as well as being a worthy target of exploration in its own right.

This report offers **Assessments** of the current state of Venus exploration, and **Actions** for the U.S. and international Venus community, as well as NASA, to consider. **This strategy is a living document, and should be updated as warranted.**

# 1 VENUS AS A SCIENCE NEXUS

[1] *Between the beginning of the 1960s and the end of the 1990s, the United States and Soviet Union between them dispatched 35 missions to Venus.*

[2] *NASA's Magellan radar mission operated in Venus orbit from 1990 to 1994.*

[3] *The ESA Venus Express mission, focused principally on the planet's atmosphere, functioned from 2005 until 2015.*

[4] *Akatsuki, or "Dawn," was the Japanese Aerospace Exploration Agency's first Venus mission, and returned data from 2015 until early 2024.*

**Venus sits at the center of some of our most important planetary science questions.** In *Origins, Worlds, and Life*, Venus as a scientific target features 261 times in the main text, and is included in 46 Strategic Research areas in seven of the 12 priority science questions—spanning the evolution of the protoplanetary disk; the origin of the inner Solar System worlds; the link between solid-body interiors and surfaces; the properties of solid-body atmospheres, exospheres, and magnetospheres; the prospect for a catastrophic loss of habitability in Venus' past because of climate change run amok; and what Venus can tell us of large, rocky worlds generally. Indeed, VERITAS, DAVINCI, and EnVision all advertised their multi-thematic relevance to a future in which Venus is a critical science destination.

Despite Venus being the target of almost three dozen missions[1] between the 1960s and 1980s, **we know remarkably little about this Earth-size world today** compared with the rest of the inner Solar System. Radar data from the Magellan orbiter mission[2] show a wealth of tectonic and volcanic landforms on Venus, but no evidence of a global plate tectonic system [1]. What those radar images do reveal are several high-standing, heavily deformed regions called tesserae, which are tempting to compare to Earth's continents. But we have neither reliable geochemical nor mineralogical data with which to determine the compositions of these regions (nor any others), and thus how they were formed. Although tantalizing clues about tesserae have been gleaned from more recent orbital missions including Venus Express[3] and Akatsuki[4], these elevated terrains—occupying around 7% of Venus by area—remain enigmatic but critical to understanding Venus' evolutionary history [2].

**We similarly lack important chemical data for the planet's interior**, leaving as a mystery the materials from which Venus accreted, how and how much water was delivered during its development, and what minerals and rocks make up its crust and mantle. No mission to Venus so far has detected an internally generated magnetic field, nor evidence for remanent magnetism in its crust, and we have precious little information about the size or state of the planet's core. The water budget of Venus' interior is also unknown, even though measurements of the atmosphere and the results of tectonic models have been generally interpreted to mean a dry interior for Venus [3,4].

The Venus atmosphere is a dense, $CO_2$-dominated greenhouse that renders the surface hotter than a self-cleaning oven and under a pressure equivalent to a depth of about 900 m underwater on Earth. The atmosphere revolves around the planet's axis up to 60 times faster than the solid body, but the reason for these vastly disparate rotation rates is unknown. There are numerous reactive trace gases present, likely many undetected especially below 20 km altitude, but we do not know when or how this atmosphere and current climate developed—is it a relict primordial atmosphere, or was Venus once far more clement and perhaps even Earth-*like*? **If not, why not? If so, when and why did Venus' climate diverge so drastically from that of Earth?**

**This understanding is crucial if we are to establish why our own planet has remained habitable almost since its formation.** *Origins, Worlds, and Life* emphasized that Venus is a remarkable science target because, as a large rocky planet with a massive atmosphere, it holds the key to understanding how and why plate tectonics, intrinsic magnetic dynamos, surface liquid water, and habitable conditions can emerge on such worlds—or how and why *not*. Because Venus is closer to the Sun than Earth or Mars, the chemistry of its interior, surface, and atmosphere holds important constraints for questions of cosmochemistry, planetary nebula processes, and the evolution of the Solar System itself. Further, **Venus serves as a cautionary tale when interpreting the limited information we have for large rocky exoplanets**: at present, we can only tell if a terrestrial exoplanet is Earth-sized, not Earth-like, and a promising exo-Earth candidate could well turn out to be an exo-Venus. Determining why Earth and Venus differ so much today will powerfully enable our efforts to distinguish Earth- from Venus-like worlds around other stars, and their divergent evolutionary histories **[5,6]**.

**Assessment 1.A**     **Venus is uniquely positioned as a science exploration target to address many high-priority, Decadal-level planetary science questions.**

## 1.1. Decadal Questions at a World Just Next Door

[5] *Venus Emissivity, Radio Science, InSAR, Topography, and Spectroscopy.*

[6] *Deep Atmosphere Venus Investigation of Noble gases, Chemistry, and Imaging.*

As for virtually all other Solar System destination, Venus' mysteries will be unveiled with robotic spacecraft exploration. The 2021 selections of the VERITAS[5] and EnVision orbiters, and the DAVINCI[6] flyby/descent probe, as well as the first-ever private mission by Rocket Lab, will begin to revolutionize our understanding of Venus just as Mars Global Surveyor and Mars Reconnaissance Orbiter did for the Red Planet in the 2000s. Yet these new Venus missions by design tackle a subset of key science questions for Venus, while programmatically reducing a variety of risks for future missions—**emphasizing that no one (or three) missions can resolve every facet of Venus science**, nor respond to the findings of the others as a new mosaic of Venus emerges.

**Much orbital-based in situ and remote-sensing science remains**, such as high-altitude monitoring of seismic activity via perturbations to airglow, or surveys that search for recent or ongoing volcanic activity via temperature change on the surface or chemical signatures in the atmosphere. Moreover, efforts to understand the planet's atmospheric superrotation and atmosphere–space environment interactions that can drive atmospheric dynamics and loss to space can be readily tackled by spacecraft in a variety of orbits, be they low- or high-altitude, inclination, or eccentricity.

For Venus' atmosphere, **we must go beyond a single vertical profile to understand its three-dimensional, time-variable cycles and interdependencies**. The dynamics of the lower and middle atmosphere are poorly understood, including the extent to which vertical mixing takes place, and even after the DAVINCI mission the Venus cloud layers, and their prospective habitability, will remain largely unexplored. We must also understand the interactions of the upper atmosphere with the space environment. Studies of other planets, including Earth and in particular Mars, have shown that such interactions, and especially atmospheric loss to space, can be fundamental in shaping the evolutionary path of a planet [7]—providing a crucial constraint if we wish to fully characterize Venus evolution.

> *we need to learn just how alive Venus is*

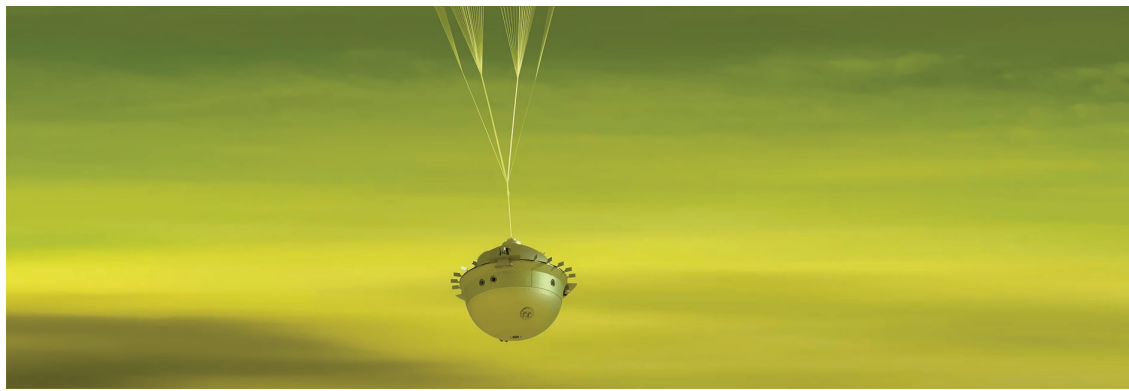

*The DAVINCI probe will fall through the Venus atmosphere, taking about an hour to descend to the surface and taking measurements as it goes. Close to the surface, the probe will take nested descent photos of the Alpha Regio tessera. Credit: NASA/GSFC*

For its surface, we need to truly determine the **nature, composition, state of weathering, and evolution of surface and near surface materials**, from the lowest-lying volcanic plains to above the emissivity "snow line" in the high-standing tesserae regions. Important here, too, is the question of the extent (if any) of surface or subsurface materials injected into the atmosphere to act as cloud condensation nuclei, ultraviolet absorbers, or even nutrients should the cloud layers be habitable. For the interior, we have to establish **the size, state, and composition of the planet's core, and**

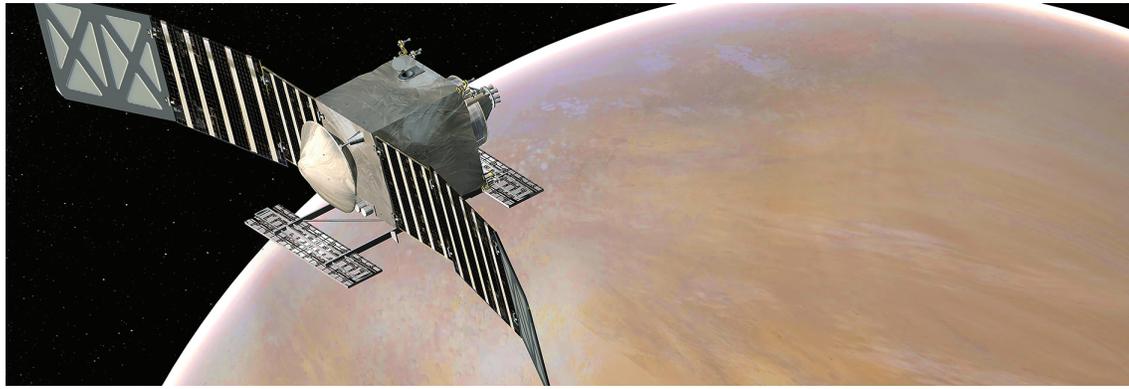

*The VERITAS orbiter will acquire global, high-resolution radar image and topographic data for Venus, together with geophysical measurements and spectral observations. Credit: NASA/JPL-Caltech*

**the structure of its mantle**. We need to learn just how alive Venus is, volcanically and tectonically, plumbing the subsurface and understanding the planetary interior.

And, although requiring of technical capabilities that are clearly still in the future, **the scientific rationale for roving on and even returning samples from Venus is every bit as applicable and compelling as it is for the Moon and Mars**—and perhaps more so, given the seeming scarcity of Venus meteorites in the world's collections. Indeed, understanding whether some exotic meteorites here on Earth are from Venus will require in situ analytical measurements.

**Orbiters are among the most mature space technologies we can deploy to Venus**, and depending on their orbit can reasonably be expected to operate for years or, by comparison with those at Mars, for decades. Yet many of our major science questions for Venus require in situ platforms with modern analytical instrumentation. **Balloon-based aerobots** can characterize the atmosphere's chemical and physical properties over much longer time scales than even the most advanced descent probes, and are ideal platforms for searching for evidence of ancient and modern magnetic fields and sources of seismicity. **Modern lander missions** can provide much better rock chemistry and definitive mineralogy than their precursors could, and one or more landers equipped with high-temperature electronics can return unprecedented, long-term science measurements of seismic activity or weather, for instance.

**Importantly, Venus' proximity to Earth confers a unique advantage** compared with almost all other targets in the Solar System where Decadal Science might be addressed. This proximity means that the launch energy to reach Venus is much less than to destinations beyond Mars—and that, once at Venus, very much more data can be communicated to Earth relative to an equivalent communication system at the outer planets. **Trip times to Venus are measured in months rather than years,** helping avoid long-duration cruise phases (and their associated costs). More importantly, scientists and engineers have the possibility of participating in *several* Venus missions in a career, even as spacecraft are dispatched to other, farther destinations.

Assessment 1.B  **Venus offers an accessible destination with more frequent launch windows and shorter cruise times—and commensurate reductions in cost and technical risk—compared with bodies closer to the Sun, beyond the frost line, or even Mars.**

Assessment 1.C  **The proximity of Venus to Earth means a much more rapid science return compared with sample return missions or those to more distant targets.**

## 1.2. A Planned Sequence of Missions to Venus

**A sequence of Venus missions is not just a theoretical possibility.** The Mars Exploration Program, initiated in the 1990s and also benefiting from trip times of less than a year, has consisted of a sequence of alternating orbital and in situ missions, each building on the last. That general approach can be replicated at Venus.

> *an integrated science program is greater than the sum of its parts*

But, as we look forward to replicating the Mars experience at Venus, there is a key difference arising from the dense Venus atmosphere. Designers of entry, descent, and landing (EDL) systems at Mars have had to deal with the conundrum of there being not enough atmosphere to be truly useful for slowing landers but too much to ignore. Many Mars landers have failed, and landing—especially of large payloads—continues to pose enormous challenges. **In contrast, aerodynamic forces at Venus can be used to passively control descent and the thick atmosphere enables sustained flight by multiple aircraft types from above the clouds to, conceivably, at the surface.**

An integrated science program is greater than the sum of its parts, as has been repeatedly shown for both lunar and Martian science. **The scientific breadth of essential and unanswered questions about Venus is staggering**, and is best addressed through the synergies created by multiple missions that build on each other through time.

Key to this approach is *planning*: although the upcoming VERITAS, DAVINCI, and EnVision missions will perform complementary science investigations, they were competitively selected and so not proposed as a single set of interrelated missions. With the existing or growing interest in Venus by nations such as Japan, India, China,

Korea, and others, the ability to establish and develop partnerships in the exploration of Venus will become ever more important.

**Assessment 1.D** The environmental conditions at Venus are demanding, but also offer unparalleled advantages to in situ exploration and mobility relative to any other inner Solar System planet beyond Earth.

**Assessment 1.E** An openly planned, integrated, and diverse set of missions and mission platforms and measurement approaches are needed to explore the most critical facets of Venus as a system, which in turn will offer new discoveries relevant to a much broader range of key planetary science issues.

**Action 1.A** **Working with the Venus community, NASA could consider how to implement a sustained, integrated program of technology development and exploration of Venus that recognizes, leverages, and values the roles that international cooperation will play to fully leverage the opportunities this planetary science nexus offers.**

# 2 EXPLORATION-ENABLING RESOURCES

Venus provides in a single world one of the most diverse, technically challenging combinations of environments available: an orbital domain that permits the use of orbiting platforms with well-suited instrumentation but challenged by a thick, opaque atmosphere; a middle atmosphere with the most Earth-like environmental conditions in the Solar System but with high variability across altitudes; and an extreme, high-pressure high-temperature setting at the surface that strains existing technical capabilities.

The exploration of Venus therefore requires an **integrated approach** that couples technology development with driving science questions and adapted measurement approaches, and vice versa.

The science questions that require comprehensive and diverse long-term landed science measurements are among the **most pressing** for Venus—but they require more technological investment. On the other hand, the technologies needed for more narrowly focused short-lived landers, aerial platforms, and especially orbiters are **more mature**, enabling compelling science at a **lower risk and cost threshold**.

## 2.1. Venus Platform Needs

**Variable-altitude, balloon-borne aerial robots with modern instruments will allow unprecedented exploration of the middle Venus atmosphere.** By changing altitude[1], such aerobots can take advantage of prevailing winds at different altitudes, providing some amount of directional control as the platform moves with the zonally super-rotating atmosphere [8]. From within the atmosphere, investigations into cloud habitability, atmospheric chemical and physical cycles, geophysics, and even surface

---

[1] *Altitude variation is achieved by pumping lifting gas from one reservoir to another, changing the balloon's volume and thus its buoyancy.*

imaging from below the cloud base become possible [9]. Key to this type of platform are advances in autonomous guidance, navigation, and control, solar power and energy storage, and directional communication systems. In addition, Earth-analog flights to demonstrate aerial deployment and inflation and to gain science operations experience in operating this entirely new type of platform are crucial. **Many of these technologies can be brought to technical readiness in the near-to-medium term.** As more is learned about Venus, we can expect this capability to evolve to include platforms able to descend below the clouds, multiple aerobots deployed from a single entry vehicle to perform networked science investigations, and even the acquisition from the surface and return to the clouds for in situ analysis of samples by multi-element aircraft.

> *long-lived, high-temperature systems would be game changing for Venus*

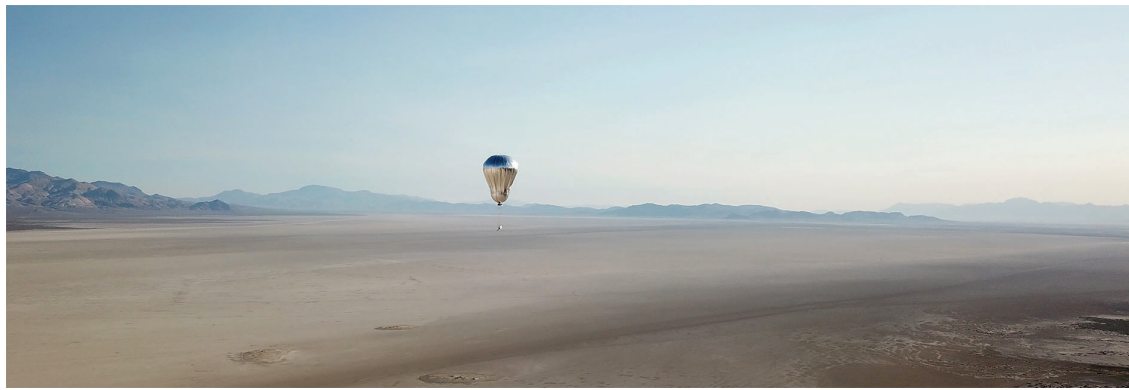

*A subscale prototype of JPL's variable-altitude aerial robot, constructed from Venus-compatible materials, undergoing flight testing over the Black Rock desert in Nevada, USA in 2022. Credit: Aerostar/NASA JPL*

**Short-term landers can provide measurements of rock chemistry, mineralogy, multispectral imagery, and even near-surface atmospheric measurements and surface–atmosphere chemical and physical interactions.** Modern short-duration landed platforms can carry mineralogical and petrological assessment instruments that would provide a multi-generational advancement beyond the Venera-era landers of the 1970s and 1980s [10]. Moreover, contemporary short-term lander designs enable lifetimes of up to eight hours of surface operations, thus allowing far greater precision (and more measurements) than those twentieth-century missions. Such missions will benefit from continued advances in thermal control to extend mission lifetime, autonomous guidance for precise targeting to landing sites—including those identified with VERITAS and EnVision imagery, and especially in the tesserae—autonomous science to optimize the use of the brief surface lifetimes, and high throughput instrumentation for in situ analysis.

**The advent of long-lived, high-temperature systems would be *game changing* for Venus.** Extended-life landers could provide the first in situ measurements of temporal phenomena at the surface such as levels and nature of seismic activity, long-term surface weather conditions across diurnal cycles, and chemical monitoring of the atmosphere and its interactions with surface materials. Advances in high-temperature technologies[2] to enable a spacecraft to operate for months at the Venus surface are promising: high-temperature sensors, communication, power storage, and other subsystems are in limited (by investment) states of development, and in some cases nearing mission viability as technology demonstrators [11]. Continued development of high-temperature technologies, including memory, power, actuation, communication, and imaging, is crucial for a sustainable Venus exploration program. Long-duration surface operations (on timescales of weeks, months, or even years) on par with conventional systems at Venus remain some way off, but even a high-temperature lander implemented as a technology demonstrator in the near-term would represent a major step forward, either as a standalone mission or as a SIMPLEx-supported secondary payload.

² *One example of high-temperature technology uses silicon-carbide-based electronics, including memory, which are resilient to the elevated Venus surface temperatures*

**Assessment 2.A** **The critical science questions that will lead to a transformational understanding of the Venus atmosphere and surface require in situ platforms that are enabled by maturing technologies.**

**Assessment 2.B** **A long-duration technology demonstrator lander deployed to Venus in the early 2030s would provide crucial insights into and further motivate the in situ exploration of the Venus surface and lower atmosphere.**

## 2.2. Venus Technology Needs

³ *Terrain-relative navigation shrank the landing ellipse of the Mars2020 Perseverance rover from 3,200 m to 40 m or better.*

Whether targeting the rolling plains or the tesserae, **terrain-relative navigation** and **hazard avoidance technologies** that have been applied so successfully to Mars would shrink landing ellipse sizes and thus enable the precision targeting of high-value science landing sites[3]. Characterization of the Venus surface at lander scales (e.g., with DAVINCI descent images, dedicated imaging dropsondes, high-resolution radar imaging from orbit), especially for the tesserae, could usefully inform the safe landing requirements for future landers. **Ultimately, surface mobility will be key**: sampling multiple sites on the Venus surface with a comprehensive instrument suite will enable the kind of transformational science now being performed at Mars by flagship-class analytical chemistry rovers.

Along with developing spacecraft component technologies, honing **instrumentation and sample handling** for Venus-specific use will lead to improvements in all exploration modalities. For example, reducing size and power requirements, yet increasing the speed and precision, of both stand-off and direct-sample landed instruments that have been used previously on Mars will lead to improvements in Venus landed science capabilities. Such improvements are needed to characterize the major, trace, and isotopic elemental abundances and bulk mineralogical, petrological, and petrographic properties of Venus surface materials sufficiently to resolve the key outstanding geochemical questions we face, as currently available instruments require environmental protection and do not yet have realistic high-temperature operational capabilities.

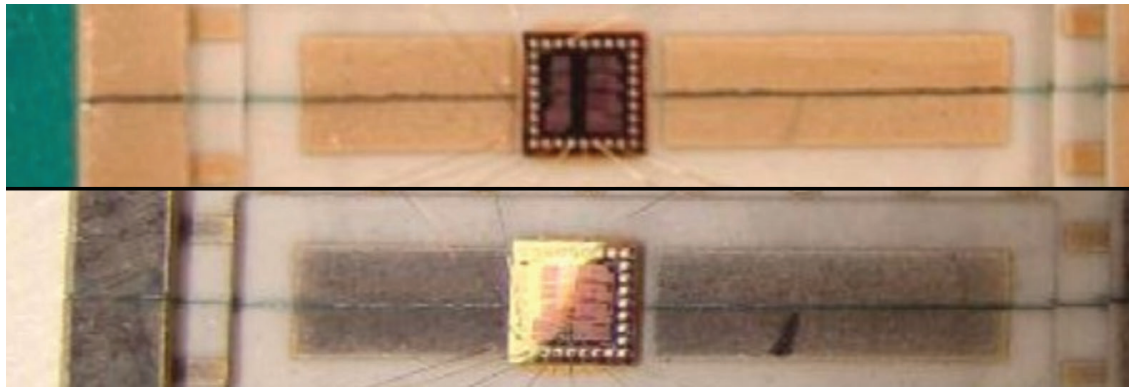

A silicon-carbide clock integrated circuit before (top) and after (bottom) being subjected to Venus surface temperature and pressure conditions for 60 days in the Glenn Extreme Environment Rig at NASA Glenn. The IC operated successfully throughout this time. Credit: NASA GRC

> "…these advances would enable a new generation of short-lived lander missions"

A **hybrid short-/long-lived lander** could take advantage of the number of high-temperature analytical systems available in the near term to provide valuable, narrowly scoped, unique, and/or complimentary time-series data to more conventional and presently available instruments, expanding the possible science return from a Venus surface platform. **All of these advances would enable a new generation of short-lived lander missions** with dramatically increased science impact compared with the 1970s and 1980s-era missions from the USSR, which were restricted to lifetimes of less than two hours, had limited ability to interact with the surface to collect samples or gather data, and were constrained by instrument capabilities of the time.

**Aerobot-borne analytical instruments will similarly benefit from advances in sample handling, throughput, and measurement precision**, as well as from high-temperature technologies. For example, the ability of an aerial platform to operate below the cloud deck, where temperatures readily exceed 100°C and approach 400°C or more within a few kilometers of the surface, will expand investigations to the entire atmospheric column on the planet, from the exosphere to the base of the troposphere.

Science and technology create a mutually beneficial cycle: **new science questions drive new technologies to answer them, and new technologies in turn motivate new science questions for us to ask**. The advancement of Venus-applicable technologies has been enabled by NASA Planetary Exploration Science Technology Office (PESTO) programs such as MatISSE[4], PICASSO[5], and HOTTech[6], resulting in important advances in surface power generation and energy storage, communications systems, data storage, and high temperature sensors. NASA Glenn Research Center has developed concepts for incorporating these technologies in surface platforms, aided by the **Glenn Extreme Environments Rig** (GEER), but further work is still needed to bring high-temperature technologies to systems and applications levels of readiness. This technical maturity could be substantively advanced through a third HOTTech solicitation.

NASA has also supported the development of component technologies and testbeds for aerial platforms through the SBIR and EPSCoR programs. The Jet Propulsion Laboratory (JPL) has developed prototype variable-altitude balloons with discretionary funding programs integrating some of the SBIR and EPSCoR work. Here again, **additional development work remains to mature these technologies**. Thus, establishing a new "CloudTech" program, focused on technology development for the upper, middle, and lower cloud layers of Venus and including aerial platform deployment, navigation, power, and communication systems, would powerfully address the needs of the next-generation in situ aerial Venus missions.

Importantly, Venus is not the only planetary body with demanding atmospheric and/or surface environments—the Mercury dayside, Titan, and the upper atmospheres of the giant planets are prime examples—so **advancing Venus technology and experimental facilities generally also facilitates future exploration elsewhere in the Solar System**, at low and high atmospheric pressures and temperatures.

All in situ science missions to Venus require entry through the atmosphere. Thermal protection systems (TPS) able to withstand this extreme entry environment are thus a critical component of such missions, and similar TPS may be necessary for implementing aerocapture at Venus. Over the past decade, NASA has invested in the development of 3D-woven TPS, including the Heatshield for Extreme Entry Environment Technology (HEEET) capability, which has been matured to TRL 6. The motivation for HEEET was to fill a TPS capability gap resulting from the loss of the carbon–phenolic TPS that enabled the Pioneer Venus mission in the 1970s. Three-dimensional-woven TPS capability is applicable to a broad range of missions including Venus, the outer planets, and high-speed sample return missions, but there are no missions at present baselined to utilize HEEET. Yet this technology utilizes a set of vendors with unique capabilities that required considerable investment to develop and certify and is implemented only by NASA, so is at high risk of loss to the community without active risk mitigation. It will be unmanageable for competitively selected, cost-capped missions to absorb the cost and schedule required to restart TPS production if the expertise and capabilities behind this technology atrophy.

---

[4] *Maturation of Instruments for Solar System Exploration.*

[5] *Planetary Instrument Concepts for the Advancement of Solar System Observations.*

[6] *NASA has solicited two Hot Operating Temperature Technology (HOTTech) calls, one in 2016 and the other in 2021. So far, 19 projects have been supported by this program.*

| Assessment 2.C | Many of the highest technology needs (e.g., aerial platforms, high-temperature systems, new analytical chemistry instruments, navigation lidar) for next-level Venus science are also technology needs for multiple other planetary targets and target environments. |
|---|---|
| Assessment 2.D | A third HOTTech call to mature high-temperature electronics for a Venus pathfinder long-duration surface mission, together with a new CloudTech program to infuse technologies to enable an aerobot mission with a lifetime of months or more in the clouds, would represent important next steps on the journey to develop truly long-lasting complete systems for the surface of Venus. |
| Assessment 2.E | 3D-woven TPS technology is an enabler for lander and aerial platform missions, but the lack of mission cadence requiring TPS of this nature threatens its availability to NASA for exploration in the 2030s and beyond. |

## 2.3. Venus Communications Needs

[7] *The Mars Relay Network currently comprises both NASA and ESA orbiters, which support data transmission from the Curiosity and Perseverance rovers.*

An indispensable facilitator of complex surface operations, including by the Mars Curiosity and Perseverance rovers, is the **satellite-based communications "proximity-relay" infrastructure** at Mars. The Mars Relay Network[7] comprises both NASA and ESA orbital assets, and is a valuable example of international cooperation in planetary exploration. At Venus, an orbital communications infrastructure would both aid aerial platform tracking and provide aerobots and landed assets—whether short- or long-duration in nature—with continuous data reception, substantially enhancing the science return of any in situ investigation.

Importantly, its proximity to Earth means that Venus is commonly used for gravity assists, thus offering more opportunities for secondary payloads than Mars or any other destination. And, by utilizing **SmallSat or even advanced CubeSat technologies for a communications relay network at Venus**, such relay satellites can likely be implemented at a relatively low cost. Innovations in SmallSat capabilities at Venus could also demonstrate independent science investigations such as mutual radio occultation for atmospheric studies as well as meter-resolution side-looking radar imaging, bridging the current gap between Magellan and upcoming orbital SAR instruments and sub-cloud descent imaging, as planned for the DAVINCI mission.

Assessment 2.F  An orbital communications infrastructure at Venus will be remarkably enabling for planned and future Venus exploration, and could be flown as a stand-alone, low-cost spacecraft or, modeled on the Mars Relay Network, could be mandated for all future Venus orbital spacecraft.

## 2.4. Venus Infrastructure Needs

The development of electronics, instruments, and spacecraft subsystems for the Venus clouds, lower atmosphere, or surface requires a **robust set of environmental facilities** in which they can be tested, validated, and optimized. Such facilities also provide unique opportunities for collateral benefits to science, from investigating how surface materials physically and chemically weather under extreme temperatures and pressures to testing different candidate aerosol species that may be present in the cloud layers. Several laboratory capabilities already exist for simulating Venus-relevant conditions, including the GEER at NASA Glenn, the JPL-Caltech Venus Cloud Simulator Facility, and the Planetary Cloud and Aerosol Research Facility at JPL, which can replicate conditions at the surface and the processes that form and sustain the planet's sulfuric acid clouds. But the questions we have in hand, the upcoming Venus missions—VERITAS, DAVINCI, EnVision, and from Rocket Lab, as well as prospective missions by India, Japan, China , Korea, and others—and the future discoveries that await us all **require that laboratory and experimental must be not only maintained but expanded**. New investments in Venus test chambers in the U.S. for DAVINCI and in Europe for Venus exploration generally could be catalysts for new, widely available Venus environment facilities and enablers of future missions to the planet.

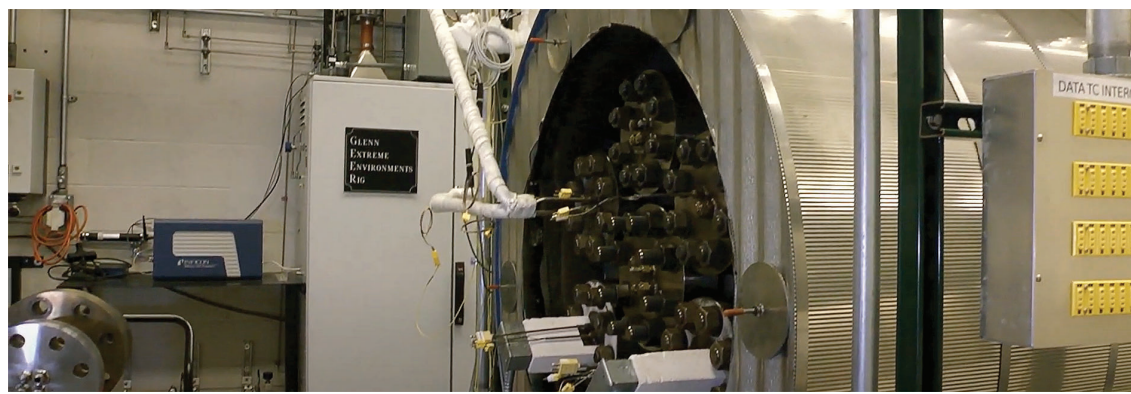

*The Glenn Extreme Environments Rig at NASA Glenn Research Center is capable of simulating atmospheric conditions for Venus, including those at the surface. Credit: NASA GRC*

There is a yet more pressing need, affecting not only Venus exploration but planetary investigations generally: **the path from component and subsystem technology to a fully integrated, flight-ready system**. Traditionally, NASA technology development is carried out at the component or instrument level, with limited opportunity for integration and testing of subsystems unless in the context of an already-selected mission or of a flagship under implementation. The result is that new paradigms for mission implementation are not likely to be selected unless risk (real and perceived) is reduced—but there is currently no viable route for demonstrating those new implementations as integrated systems to reduce such risk. Venus lander and aerial platform concepts will benefit hugely from an approach that integrates component and subsystem technology to demonstrate basic systems-level viability and appropriate refinement and optimization, as was accomplished for the Mars Exploration Program with efforts including (but not limited to) the Mars Instrument Design and Development Program in the 2000s.

**Assessment 2.G**  **Coordinated laboratory facilities and instrument development, supporting both technology and science objectives, are critical to the success of future Venus missions and together represent the foundation for Venus exploration infrastructure at a "Program" level.**

**Action 2.A**  **NASA could reinforce its approach to Venus exploration by identifying essential technologies required to address key Decadal-level science questions and supporting the development of those technologies from the component to the systems level, including but not limited to aerial vehicles, high-temperature electronics, autonomous navigation and sampling handling, and thermal protection systems.**

**Action 2.B**  **NASA could strengthen its planned and future investments in Venus missions by exploring all options to establish an on-orbit relay communications asset at the second planet.**

# 3 BUILDING EXPLORATION PARTNERSHIPS

As a unique planetary science nexus, Venus offers considerable value to researchers beyond those studying the planet's interior, surface, and atmosphere. Venus holds lessons for those investigating topics as diverse as planetary habitability, cosmochemistry, aeronomy, the space environment, heliophysics, exoplanets, and even some creative aspects of human exploration. What is vital, then, is demonstrating that Venus is the ideal laboratory for the study of these cross-objective and cross-disciplinary topics. Similarly, there is growing interest internationally and in the commercial sector in the exploration of Venus. Thus, by partnering with as broad a cohort of stakeholders as possible, within and beyond the United States and the national scientific community, the cause of Venus exploration is strengthened.

## 3.1. Partnerships with Other Scientific Groups

### 3.1.1. Space Physics and Heliophysics at Venus

The Mars Express and MAVEN[1] missions have enabled the growth of a large and active Mars **upper-atmosphere community**, which has shown that a full understanding of the dynamics and evolution of the planet as a whole requires comprehensively characterizing the transport of materials through that atmosphere. To what extent do lower-atmosphere processes such as volcanic activity affect the atmosphere? What roles do the solar cycle and space weather play?

With the upper-atmosphere and space environment insights we have for Earth, **Venus offers a crucial third piece of this comparative planetary puzzle**. Although Pioneer Venus Orbiter and Venus Express made great strides in some aspects of understanding the planet's atmospheric dynamics and chemistry, they were not designed to investigate these processes as comprehensively as has been done at

---

[1] *Mars Atmosphere and Volatile EvolutioN.*

Mars. A Venus mission with a focus on upper-atmosphere–space environment interactions is of major interest to the planetary atmospheres community generally, encompassing those scientists who have yet to focus on Venus.

The heliophysics community has a vested interest in long-term monitoring of the Sun, but currently lacks any such means at Venus. A mission to the second planet capable of remote sensing of the Sun or of in situ measurements of the solar wind will, in addition to Venus science, **enable studies of stellar processes and solar wind evolution through the Solar System**, as well as offering valuable solar monitoring.

### 3.1.2. *Venus as an Early-Earth Analogue*

Venus' tectonic and volcanic characteristics and evolution could provide clues to those Earth scientists studying the **geochemical, geophysical, and geodynamic properties of the Hadean and Archean**, the geological record of which is now largely lost to us through subduction and erosion. For example, Venus' enigmatic coronae show features that resemble small-scale convergent tectonic plate boundaries on Earth [12]. Venus' highly deformed terrains, the tesserae, have long been regarded as potential counterparts to the continents on Earth, perhaps even dating to a time of sustained habitable conditions at the surface that included liquid-water oceans [13]. The sedimentary cycle on Venus remains poorly understood [14], but by comparison with Archean and Proterozoic rocks may offer clues to how its environment changed over time, perhaps even from a habitable period to that of today. Missions that offer new insights into corona formation, tessera composition, history, driving geodynamic regime, and sedimentary processes will impact not only how we interpret Venus' geological and climate evolution, but our understanding of the early processes that shaped our own world.

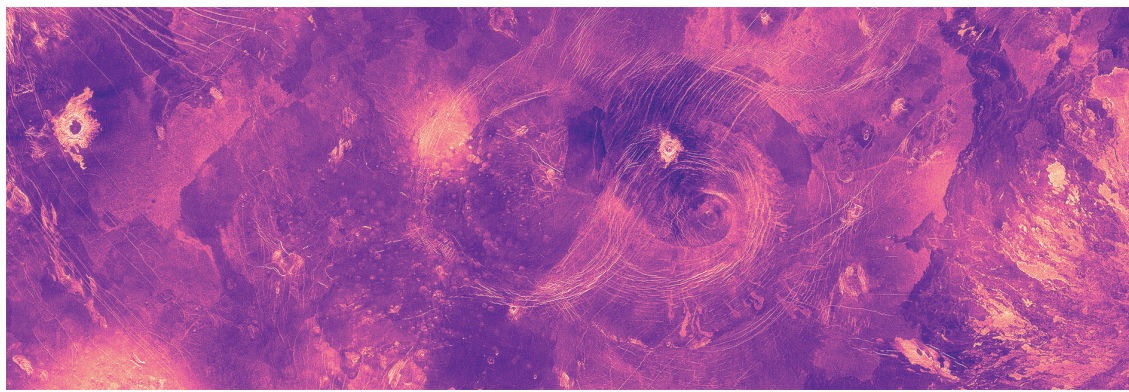

*Venus abounds with coronae—volcanotectonic structures characterized by annular fractures and thought to result from mantle upwellings. The thin mechanical lithosphere of Venus, in which these landforms develop, may be similar to that of Archean Earth. Credit: NASA/Paul Byrne*

### 3.1.3. *The Exoplanet Next Door*

The field of extrasolar planets is flourishing, with more than 5,500 confirmed exoplanets found to date and an increasing ability to not only detect but characterize these worlds. **A major focus of exoplanet research is the search for Earth-*like* planets**, thought to be a subset of so-called "eta-Earths"—those Earth-size or larger rocky worlds within a star's optimistic habitable zone [e.g., 15]. Vital to this search is an understanding of the likely occurrence rates of "eta-Venus" worlds, and whether such worlds might be the more common type to develop, which **requires establishing why the climate histories of Earth and Venus have so drastically diverged**. Engaging with the exoplanetary science community is therefore an obvious and important step in advancing the study of Venus—and vice versa.

² *The James Webb Space Telescope, in operation at the Sun–Earth $L_2$ Lagrange point since early 2022.*

³ *The Habitable Worlds Observatory, designed to search for Earth-like worlds and scheduled to fly in the 2040s.*

Relatedly, many exoplanets likely to be studied by JWST[2] and HWO[3] are expected to be tidally locked. Since it is the slowest-rotating terrestrial planet in the Solar System, Venus may be the best place available to us to study how the lower, middle, and upper portions of a rocky planet's atmosphere respond to slow planetary rotation.

### 3.1.4. *The Moon to Mars—Via Venus*

As the images from the Parker Solar Probe's Venus flybys have shown, via the discovery of a **previously unknown window in the Venus atmosphere enabling imaging of surface emission**, data acquisition at Venus by future missions that use the second planet for gravitational assistance offers opportunistic science and represents an important step toward building partnerships across scientific communities.

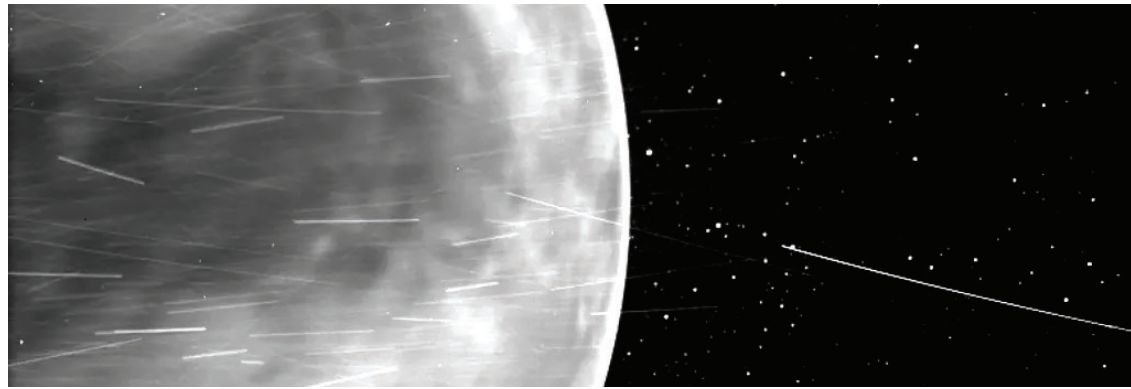

*Surface features on Venus imaged at <0.8 μm by the WISPR (Wide-field Imager for Solar Probe) instrument onboard Parker Solar Probe. The dark region at left is part of the Aphrodite Terra highland, which is topographically elevated and is thus cooler than the surrounding plains. Credit: NASA/APL/NRL*

Venus could also serve as an important stepping stone for *human* **exploration of the Solar System**. Venus can be—arguably should be—how we travel to and from Mars. Crewed flybys of Venus on the outbound or return leg to minimize delta-*V*, entry velocities for crewed spacecraft, and mission duration are valuable precursor missions for human spaceflight to Mars. These flybys further offer abort-and-return-to-Earth options as well as opportunistic science during closest flyby approach. Incorporating Venus flybys in crewed Mars missions turns human exploration of merely one destination into **human exploration of the Solar System**, changing our aspirations from simply boots in the red dust to a much grander endeavor—and, uniquely, de-risking

> *if… the planet were once legitimately Earth-like, the astro-biological importance of Venus would be unparalleled*

human spaceflight to Mars and offering a testbed for human exploration technologies. Gravity-assist maneuvers performed by non-crewed missions also offer an exciting opportunity to acquire more data at Venus.

*Origins, Worlds, and Life* placed a great emphasis on process, not planet—and **Venus is that rare Solar System body where questions that transcend and cross-cut multiple distinct disciplines can be answered**, and fed forward across the driving themes in NASA's science portfolio. Forging new and developing existing relationships with communities such as planetary atmospheres, heliophysics, Earth science, exoplanetary, and human exploration through research collaborations, conference sessions, workshops, and special journal issues, will serve to further catalyze Venus exploration.

### 3.1.5. *Venus as an Astrobiology Destination*

Although conditions in the upper clouds at Venus are **more clement than virtually any other destination beyond Earth**, the actual state of habitability of those clouds remains unknown. The detection from Earth of phosphine in the Venus atmosphere remains contested [16,17], and the chemical processes that *could* sustain life there in the present are far from fully understood—even though the Venus atmosphere has been thought of as a potential abode of life for decades. Nevertheless, the field of astrobiology has grown immensely in the past two decades, and there are **exciting new opportunities** to marry that collective expertise and work with the new insights into prospective habitability of Venus' atmosphere. The Rocket Lab mission is one such example of the application to the second planet of perspectives that heretofore have largely focused on Mars and the outer Solar System moons. Moreover, if VERITAS, DAVINCI, EnVision, and future missions were to establish that the planet were once legitimately Earth-like, the astrobiological importance of Venus would be **unparalleled**.

Assessment 3.A — **The exploration of Venus offers substantial value to the wider planetary, heliophysics, and Earth science communities, and there is major scope to develop and grow new, mutually beneficial partnerships with diverse scientific disciplines.**

## 3.2. International Partnerships

**Space and planetary exploration has long been an internationally collaborative affair**, evinced by for instance the International Space Station, the James Webb Space Telescope, and the Mars Sample Return project. Venus exploration is no different:

VERITAS will fly several contributed elements from ESA member states, and EnVision will carry a NASA-furnished radar. International participation in Venus exploration is thus already widespread, **and can only be expected to grow**.

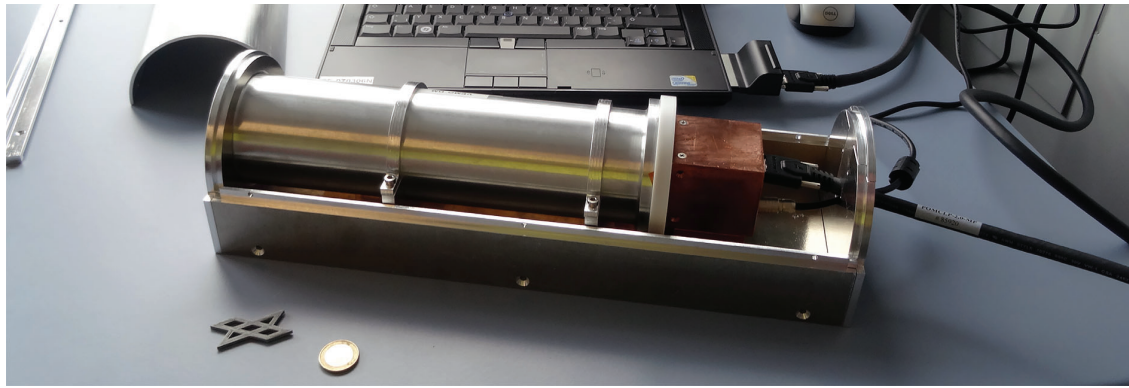

*The laboratory prototype of the Venus Emissivity Mapper (VEM) instrument, which will be carried on the VERITAS spacecraft and is provided by DLR. This instrument will return spectral data for the major terrains on Venus. Credit: DLR/Helbert et al. (2018)*

NASA's most important international partner at present is arguably **ESA**, although collaborations with individual member states' aerospace centers are crucial, too. For instance, a field campaign to Iceland in 2023 to ground-truth radar and spectral data from the VERITAS mission was co-organized by the **German Aerospace Center DLR** and JPL, and the French national space agency **CNES** has been a key participant in numerous Venus mission concepts and studies. ESA itself has had its own dedicated mission to the planet, Venus Express (which featured U.S.-based co-investigators). The **Japan Aerospace Exploration Agency**, too, has an established track record in Venus exploration with its Akatsuki spacecraft, and the agency has aspirations to return to the planet with a follow-on mission. **India** has announced plans to send its own Venus Orbiter Mission to the planet, and **China** recently unveiled an ambitious program of planetary exploration that includes a possible Venus atmospheric sample return mission. The newly formed **Spanish Space Agency** and **Korea AeroSpace Administration** are examples of other prospective partners in NASA's exploration of Venus.

[4] *The VeSCoor group was established in 2023 and, at time of writing, comprises selected members appointed by both ESA and NASA, as well as Venus mission representatives and observers from NASA, ESA, and from Japan and South Korea.*

There already exists a mechanism of international collaboration between NASA and ESA in the form of the **Venus Science Coordination group**[4] (VeSCoor), which has a focus on synergistic science activities between the VERITAS, DAVINCI, and EnVision missions. NASA's international Venus partnerships could be substantially strengthened by an entity similar to the International Lunar Exploration Working Group (ILEWG) and the International Mars Exploration Working Group (IMEWG), both of which were founded in the 1990s as fora for the exchange of ideas and strategies for the exploration of these bodies. Importantly, such a working group for Venus could feature executive members from NASA itself, empowered to engage directly with other agencies, nations, and even commercial partners to develop collaborative plans for the exploration of Venus in the 2030s and beyond. An **International Venus Exploration**

Working Group would be an appropriate entity to establish, for example, international standards for communications and data relay in Venus orbit.

Assessment 3.B    Efforts to prepare both for the data return from the VERITAS, DAVINCI, and EnVision missions and for the next generation of Venus missions will be most effective with NASA, its international partners, and the Venus community acting collaboratively and in concert, for example via groups such as VeSCoor.

Assessment 3.C    The substantial interest in Venus exploration by other national space agencies and commercial entities offers an exciting opportunity to develop and take advantage of partnerships through the establishment of an International Venus Exploration Working Group.

Action 3.A    **The Venus community should actively engage with as broad a cohort of scientific disciplines and stakeholders with a shared interest in exploring Venus as possible, to identify how resources can be pooled to do compelling, impactful, and relevant science at the second planet.**

# 4 SUPPORTING VENUS SCIENCE IN THE U.S.

The VERITAS, DAVINCI, and EnVision missions will herald an upcoming era of Venus exploration referred to as the "Decade of Venus," with the private Rocket Lab mission serving as an exciting prologue. But the realities of budgets, formulation, and implementation mean that none of the NASA/ESA missions will launch until the early 2030s, and the major science payoffs are ten or more years away as of this writing—such that we've embarked upon the "*Decades* of Venus." The fact is that these larger missions' discoveries will not begin impacting the trajectory of planetary science until the 2030s.

Yet that timing affords us the opportunity now to increase our engagement with Venus science in the United States, **maximizing the return on these upcoming missions** and the new discoveries they will inevitably make, and laying the groundwork for identifying and addressing the resulting major questions we must tackle.

Equally, continued laboratory simulations of the planet's atmosphere and surface processes, the development of new and better models of Venus' climate and interior, and sustained analysis of existing Venus data sets—from surface investigations with Magellan data to cloud dynamics measurements from Akatsuki to interactions between the upper atmosphere and the space environment with Pioneer Venus, Venus Express, and even Galileo and Parker Solar Probe flyby observations—will together best position us to maximize the science return of the upcoming Venus missions, and set the stage for more to follow.

Much Venus research in the U.S. is funded by NASA research and analysis (R&A) programs. **Recent changes to those R&A programs have been more encouraging for Venus science**, such as the inclusion of Magellan synthetic-aperture radar (SAR) mission data into the Discovery Data Analysis Program, and the recognition of quadrangle geological maps as part of the Planetary Data Archiving, Restoration, and Tools program. A catalytic "Venus Fundamental Research" initiative could help promote and apply advances in the science that will come from the missions to be launched in the early 2030s, an approach that was successful in the early years of the Mars Exploration Program[1].

---

[1] *Running from 2005 through 2013, NASA's Mars Fundamental Research Program solicited research proposals focused on the geology, climate, and atmosphere of the planet with the goal of motivating future spacecraft exploration of Mars.*

| Assessment 4.A | The establishment of a fundamental research program for Venus science would place a firm focus on systems-level science questions to be tackled ahead of and in response to findings from the upcoming missions to Venus. |

**Yet there is more NASA can do to support and grow the U.S.-based Venus science community and its international partners.** For instance, there is a compelling need to finish the remaining Magellan quadrangle maps ahead of the VERITAS and EnVision missions, to enable comparisons with geological maps prepared with those future, high-resolution radar imaging and topographic datasets and to help guide selections of EnVision regions of interest. Such mapping could be an explicit part of the annual PDART solicitation. Venus atmospheric circulation and chemical modeling has been dramatically enhanced by the results of the Akatsuki mission, but much of that expertise lies outside the United States—and **so there is need to grow this skillset domestically**. Supporting the use and maintenance of existing Venus experimental facilities, and standing up new such community infrastructure as needed, is vital to fully understand atmospheric and surface chemical and physical processes at Venus-relevant temperatures and pressures—and to allow more effective communication with non-U.S. scientists. Crucial, too, is that those data are collated in appropriate archives, such as NASA's Planetary Data System, so that they may be readily and freely accessible by the U.S. and international planetary science community.

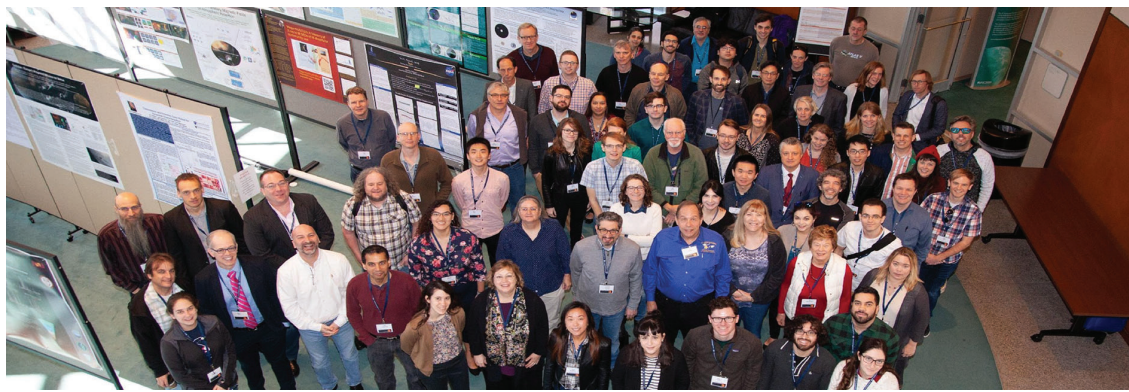

*A photograph of attendees at the inaugural Exoplanets in Our Backyard meeting, held at the Lunar and Planetary Institute in Houston, TX in February 2020. Source: https://doi.org/10.3847/25c2cfeb.8e6c355a.*

Enabling more community meetings in which scientists who study planetary processes relevant to both Venus and to other worlds **can meet and learn from each other**—such as the joint VEXAG/OPAG/ExoPAG/MEPAG/MExAG "Exoplanets in Our Backyard" workshop series[2]—will be hugely effective in demonstrating the value of Venus as a key science nexus to the broader planetary community. And increasing and sustaining internship, scholarship, and research opportunities for students and early-career scientists will be vital to ensure the successful growth of the Venus community, especially as we look to the decades ahead. Similarly, the substantial (and growing) volume of literature synthesizing data from the Magellan, Pioneer Venus, Venus Express

---

[2] This meeting series has proven highly successful at enabling joint Solar System–exoplanet research projects and collaborations; at time of writing, there have been three EiOB workshops.

Missions and others, laboratory and numerical modeling, and analogue fieldwork would be effectively complemented by **a set of new, comprehensive review articles** that encompasses planetary geoscientists, atmospheric scientists, astronomers, Earth scientists, and other constituencies with an interest in exploring Venus. These review articles would summarize in shorter form than a book the current state of knowledge of Venus science, outline the opportunities and challenges that lie ahead, and **clearly articulate the value of continued Venus exploration for astronomy, Earth, and planetary sciences generally**.

**Assessment 4.B** — The timing of the next crop of Venus missions offers a unique opportunity *now* to ready the planetary community for the science to come, by sustaining and enhancing R&A science programs, mapping and modeling efforts, coordinated review publications, and community workshops, and by ensuring that experimental and laboratory results are collated in suitable, freely accessible public archives.

**Action 4.A** — **NASA could support R&A programs and meetings to enable the U.S. and international planetary community to maximize their readiness for upcoming Venus missions and advance the scientific rationales for the generations of missions to follow.**

# 5 THE NEXT GENERATION OF VENUS MISSIONS

## 5.1. Next Steps for Directed Venus Missions

In preparation for *Origins, Worlds, and Life*, a mission concept study for a Venus Flagship Mission (VFM) was carried out in 2020 as part of NASA's Planetary Mission Concept Studies call [10]. Because this ambitious concept predated the 2021 selections of VERITAS, DAVINCI, and EnVision, it addressed a broad range of science goals that, in part, happened to overlap with those missions. Moreover, the VFM concept included a lander that would target a narrow landing ellipse in the tesserae, necessitating considerable autonomy and flight-control technology development. Ultimately, the combination of an independent costing estimate and the technical risk of landing in the tesserae and sample handling resulted in the VFM being ranked lower in priority than other, competing flagship-class concepts.

Nonetheless, *Origins, Worlds, and Life* found that a **landed flagship mission to the plains could achieve science of similar impact to the 2020 VFM concept**. A plains-targeted flagship would advance key, outstanding science questions we have for Venus without obviating the need for (and perhaps even encouraging) later landed missions to the tesserae, but would be achievable with near-term technology and could be safely designed with existing Magellan data. In turn, a flagship-class tessera lander would be enabled by the high-resolution radar image and topography data VERITAS and EnVision will return in the 2030s, and by the subcloud image and topographic data of Alpha Regio to be acquired by DAVINCI during its descent to that tessera exposure.

Importantly, a flagship-scale Venus mission was ranked highly in the preceding 2013–2022 Decadal Survey *Vision & Voyages*[1]. This concept, focusing on investigating the planet's climate and comprising an aerobot, dropsondes, and carrier spacecraft, was prioritized after a Mars sample caching mission (which became Mars2020 Perseverance), a Europa-focused mission (which became Europa Clipper), and a Uranus orbiter and probe (UOP, prioritized in *OWL* as the next flagship), and jointly prioritized with an Enceladus mission (which was ranked as the flagship to follow the

---

[1] This flagship, termed the "Venus Climate Mission," was the top and sole priority of the Inner Planets panel; an independent cost estimate placed it total price at $2.4 billion in FY2015 dollars, second cheapest only to an Enceladus orbiter mission.

Uranus mission in *OWL*). Following this ranking sequence, then, **a mission to Venus would be the natural choice for the prioritized flagship in the** *next* **Decadal Survey**.

The primacy of a Uranus flagship was secured by two major studies, the *Ice Giants Pre-Decadal Study Final Report* led by JPL and published in 2017, and the Applied Physics Laboratory's 2021 *Uranus Orbiter and Probe Planetary Mission Concept Study* for the *OWL* Decadal Survey. These studies helped identify and mitigate technical, cost, and schedule risks for such a concept, making the UOP mission a straightforward choice for NASA's next planetary flagship after Europa Clipper. Such **pre-decadal studies are therefore crucial to maturing a flagship mission concept** to the point where it is a credible choice for a decadal survey steering committee.

**Assessment 5.A** — **A lander capable of sophisticated geochemical measurements, whether directed to the plains or the tessera, and including of an experimental payload capable of an extended surface lifetime, could be a competitive candidate for the flagship mission to follow Enceladus OrbiLander.**

**Assessment 5.B** — **A comprehensive study of such a future Venus lander would position this concept as a competitive candidate for the flagship mission to follow the Enceladus OrbiLander.**

## 5.2. Next Steps for Competed Venus Missions

[2] *Escape and Plasma Acceleration and Dynamics Explorers.*

NASA's competitive planetary mission programs include SIMPLEx, Discovery, and New Frontiers. The low cost cap of SIMPLEx missions, which are funded from the Discovery program budget, requires that spacecraft are co-manifested for launch on other missions. No Venus SIMPLEx proposal has yet been selected, but the ESCAPADE[2] twin-spacecraft mission to Mars demonstrates **the scale and capability of the science investigations that could be accomplished at Venus at this price point**.

The Discovery program supports both the VERITAS and DAVINCI missions, as well as the VenSAR instrument on EnVision. The next Discovery Announcement of Opportunity (AO) is not currently expected until 2027, and the selection of a sole Venus mission concept is unlikely on the grounds of programmatic balance—unless NASA were to simultaneously pick a second mission to a different target. Nevertheless, **VERITAS and DAVINCI have faced substantial fiscal constraints** arising from broader issues in the NASA budget, with both missions now expected to launch several years after their originally proposed schedules (and with the formulation of VERITAS essentially being suspended for over a year). It is *vital* that these missions launch as expeditiously as possible—the new era of Venus science requires it.

Assessment 5.C    NASA risks undermining its current investment in Venus science exploration unless the VERITAS and DAVINCI missions launch in the planned 2031–2032 timeframe.

NASA's New Frontiers competition has, since its inception, featured a "Venus In Situ Explorer" (VISE) mission theme, focused on **addressing science questions from within the atmosphere or on the surface**. There have been multiple proposals to that theme in consecutive New Frontiers (NF) competitions and, although several were deemed fully selectable, none has so far been successful. In the run-up to the NF5 competition in 2022, NASA deleted the VISE theme from the list of permitted mission themes ostensibly for programmatic balance, despite there being no suggestion from the community nor any advisory group to do so. The next NF competition has been delayed several years, prompting a reconsideration of that deletion. The NF cost cap enables scientific capabilities that are simply not possible under the Discovery program, such as in situ exploration, and is thus **an essential part of the future of Venus science exploration**.

*" the programmatic balance argument for removing Venus from NF5 is flawed "*

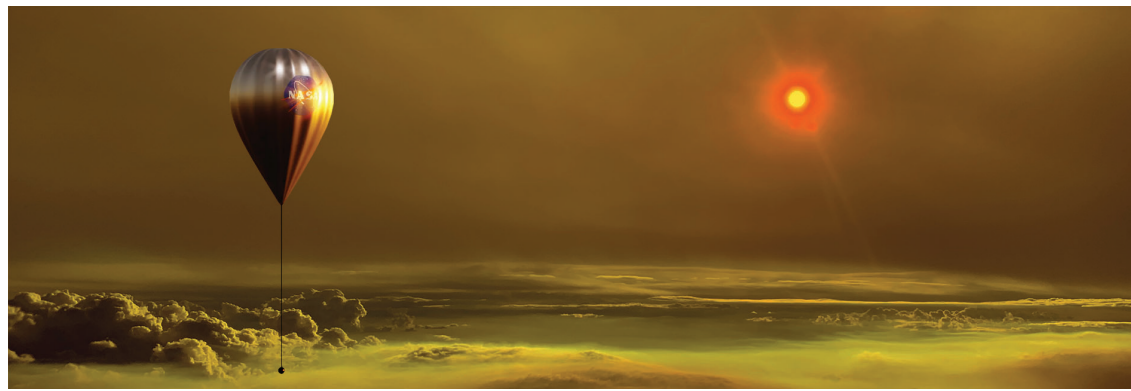

*One form of the Venus In Situ Explorer concept involves an aerobot taking physical and chemical measurements of the planet's cloud layers. Credit: Paul Byrne; balloon by Tibor Balint*

The programmatic balance argument for removing Venus from NF5 is flawed. Not only has there been a dearth of U.S. missions to Venus since Magellan ended in 1994, but there is precedent for having two Discovery and one New Frontiers missions within a decade to the same target class: the NF asteroid sample-return mission OSIRIS-REx was selected in May 2011 and the Discovery asteroid missions Lucy and Psyche were announced as the winners of their AO in January 2017.

Moreover, **data from VERITAS, DAVINCI or EnVision are not prerequisite for implementing a NF-class VISE mission**, as demonstrated by the ADVENTS mission concept carried out for *Origins, Worlds, and Life*. Indeed, the revised "VISE" theme in *OWL* explicitly lists science objectives separate to and not predicated on precursor science from those planned missions.

Assessment 5.D  The building momentum of Venus exploration will be severely inhibited unless NASA permits the community to propose to the "VISE" theme in the next New Frontiers AO.

## 5.3. Positioning for Future Mission Opportunities

Aerial missions employing fixed- or variable-altitude balloons operating in the cloud layers can be implemented with technologies that are **rapidly maturing and will be ready for the next NF and Discovery AOs**. Concepts such as ADVENTS are capable of performing studies of atmospheric dynamics and chemistry, geophysics, and even astrobiology within the NF cost cap. Importantly, however, **aerial platform technology is scalable**—and can be applied to networked science missions, subcloud vehicles, and even as part of a capstone surface sample analysis and/or return mission in the coming decades.

Moreover, when high-temperature technologies are available at the systems level, they can be paired with buoyant mobile platforms to conduct science investigations from below the clouds all the way to the surface, and would complement long-lived stationary surface assets. Because of the planet's uniquely thick $CO_2$ atmosphere, **aerobots will be the natural modality for conducting mobile surface science**, performing the same role as the Dragonfly rotorcraft will at Titan[3].

[3] *Dragonfly is an octocopter/lander, capable of traversing tens of kilometers across Titan's surface in successive flights. Dragonfly is NASA's fourth New Frontiers mission.*

All of these capabilities can be applied within the context of Venus as a nearby destination that can be reached quickly and often—underscoring the fact **that a Venus Exploration Program could be implemented with (and would benefit from) more missions that are relatively inexpensive** than fewer investigations that are each more costly. Even Venus flagships need not be necessarily as pricey as their counterparts to the outer Solar System, say, and **a Venus Exploration Program could readily include directed missions that are *not* flagship-class**, such as the New Frontiers-scale Endurance-A concept proposed for the Moon.

Action 5.A  **NASA and the Venus community should coordinate to study a new flagship mission to the plains as soon as possible, to establish the cost and technical feasibility of such a mission as the next Decadal Survey-recommended flagship.**

Action 5.B  **The Venus community should maximally leverage NASA to fly the VERITAS and DAVINCI as soon as budgetary realities allow.**

**Action 5.C**  The revised "Venus In Situ Explorer" (VISE) theme described in *OWL* should be restored to the target list for the next New Frontiers call and retained in future NF AOs until selected.

**Action 5.D**  With the delay to both missions' launch readiness dates, NASA could consider soliciting Venus-focused SIMPLEx proposals for VERITAS and DAVINCI.

# 6  A NEW VISION FOR VENUS

Venus has long been a compelling exploration target—and the 2021 selections of VERITAS, DAVINCI, and EnVision and the Rocket Lab probe have further cemented that status. The Venus community has markedly increased in size over the last few years, and we can only expect interest in the second planet to continue to grow.

Venus offers an unrivalled, nearby, and readily accessible destination for answering some of the most fundamental and pressing planetary science questions we can ask. The timeliness of Venus missions with on-target science in six to twelve months after launch makes it **a quick return-on-investment destination that crosses science community boundaries**, such that technological and scientific feed forward can happen at a rapid cadence, and that low-cost missions to the planet are possible. Moreover, the necessarily long cruise times to most other planetary destinations mean that even missions started later this decade will not return data until the late 2030s or even the 2040s—raising the prospect of **a huge data void in the next decade that missions to Venus following VERITAS, DAVINCI, and EnVision can fill**.

Many of the assessments and recommendations in this report can be effected by the Venus community, and VEXAG with its component Study Analysis Workgroups is the appropriate entity to oversee and coordinate this work—but **truly implementing these plans requires buy-in from NASA**. For example, mission concept studies necessitate that funds be directed to (and the studies be carried out by) NASA Centers, the establishment of an International Venus Exploration Working Group needs to be led by NASA HQ, and most importantly the enacting of a science exploration strategy requires a budget. **Together, these and the other actions recommended here warrant the establishment of a Venus Exploration Program (VEP).**

**The rationale for this Program already exists.** There are Venus missions in development both by national/transnational space agencies and the private sector. Other nations have indicated their intent to send spacecraft to Venus, and the international Venus community is growing. The technologies required for in situ Venus exploration are maturing. And early steps such as establishing common relay communication protocols and identifying collaborative opportunities with partner

> *the enacting of a science exploration strategy requires a budget*

space agencies can be accomplished by NASA relatively quickly and at relatively little cost, organized through, for instance, a new Venus Exploration Coordination Office.

**A new Venus Exploration Program at NASA would be represented by a budgetary line item that enables a series of missions to be conducted without each having to be independently authorized.** Securing a separate budget line demands that the Venus community clearly advocates for and demonstrates that such a Program will be affordable and consistent with NASA's overall planetary science goals and objectives. This demonstration must go beyond the planetary community and other relevant scientific stakeholders to include NASA leadership, Congress, and the public at large.

Advocating for a VEP to NASA, Congress, and the broader planetary community will be all the more effective by articulating a **clearly defined end goal**—such as a capstone surface sample return mission to be realized in several decades' time and that takes inspiration from the recent VISTA[1] mission concept study [18], but which starts with the results of the VERITAS, DAVINCI, and EnVision missions.

Previous successful exploration efforts have hinged on a clear, concise message—think *follow the water* for Mars. Quickly establishing the **"unique selling point"** of Venus, distilling it to a succinct statement, and strategizing how to advertise that statement to relevant audiences will be important in bringing together the disparate partners needed to support a Venus Exploration Program. Here, **a VEXAG public engagement plan**, developed in partnership with NASA, is important. Such a slogan could focus on the similarities and differences between Earth and Venus, the value of Venus in understand large, rocky exoplanets, or the prospect for Venus once being an ocean world.

**The 2033–2042 Planetary Science Decadal Survey is a natural vehicle through which a community-supported recommendation that NASA establish a Venus Exploration Program can be made.** The next survey membership will presumably be constituted in the late 2020s, and the mid-term evaluation of *Origins, Worlds, and Life* will take place even sooner. Thus the time to start actively pursuing a Venus Exploration Program is now.

---

[1] *The Venus In Situ Sample Transfer and Analysis mission concept was the subject of a Keck Institute for Space Studies report in 2022, and featured a comprehensive aerial platform laboratory to which samples from the surface and atmosphere would be supplied by Sample Landers and Ascent Vehicles. VISTA would be a multi-element flagship-class mission.*

| | |
|---|---|
| **Assessment 6.A** | NASA's long-term exploration of Venus is best achieved through a formal VEP. |
| **Assessment 6.B** | Identifying and marketing a compelling statement that motivates continued Venus exploration will be a potent tool for advocating for a Venus Exploration Program. |
| **Action 6.A** | The Venus community should actively work to convey the scientific importance, value, and need for a clearly defined VEP to as broad a set of stakeholders as possible between now and the start of the next planetary science Decadal Survey. |
| **Action 6.B** | The work of establishing a dedicated Venus Exploration Program should start now. |

# A     SELECTED BIBLIOGRAPHY

# B  REPORT PREPARATION

## B.1. Report Writing

The development of this report began with the establishment of a VEXAG Study Analysis Workgroup (SAW) in January 2023. This SAW quickly identified the main elements to focus on for the report, and solicited community feedback in a town hall at the 2023 Lunar and Planetary Science Conference. Writing began that summer, with a full report draft document prepared in fall 2023. After targeted feedback from specific members of the Venus community, that draft strategy was presented at the VEXAG annual meeting in Albuquerque, NM in October 2023. The SAW continued to seek community feedback at the 2024 LPSC and throughout the summer; the penultimate strategy document was shared at the 2024 COSPAR 2024 Scientific Assembly in Busan, Korea in July 2024 to engage international partners. The final exploration strategy was unveiled at the VEXAG annual meeting in Louisville, KY in November 2024, and made available online later that month.

The members of the VEXAG Strategic Plan SAW were:

James Cutts
Chuanfei Dong
Christopher Fowler
Tracy Gregg
Anna Gülcher
Gary Hunter
María Regina Apodaca Moreno
Cadence Brea Payne
Emilie Royer
Alison Santos
Paul Byrne (Lead)

## B.2. Acknowledgements

This report benefitted enormously from the feedback, suggestions, and comments from a great many people, including those from across the planetary science community, attendees at the 2023 and 2024 LPSC town halls and the VEXAG annual meetings, and scientists from several NASA centers and NASA HQ.

We particularly thank the following for their efforts in bringing this report to completion (listed in alphabetical order):

Jeffrey Balcerski
Debra Buczkowski
Michael Chaffin
Anthony Freeman
James Garvin
Lori Glaze
Noam Izenberg
Scott King
Daniel Nunes
Aoife O'Halloran
Jason Rabinovitch
Suzanne Smrekar
Bradley Thomson
Ethriaj Venkatapathy
John Vistica

# B.3. List of Acronyms

| | |
|---|---|
| ADVENTS | Assessment and Discovery of Venus' past Evolution and Near-Term climatic and geophysical State |
| AO | Announcement of Opportunity |
| CNES | Centre national d'études spatiales (French national space agency) |
| COSPAR | Committee on Space Research |
| DAVINCI | Deep Atmosphere Venus Investigation of Noble gases, Chemistry, and Imaging |
| DLR | Deutsches Zentrum für Luft- und Raumfahrt (German national center for aerospace) |
| EDL | Entry, Descent, and Landing |
| EPSCoR | Established Program to Stimulate Competitive Research |
| ESCAPADE | Escape and Plasma Acceleration and Dynamics Explorers |
| ExoPAG | Exoplanet Exploration Program Analysis Group |
| GEER | Glenn Extreme Environment Rig |
| GRC | NASA Glenn Research Center |
| HEEET | Heatshield for Extreme Entry Environment Technology |
| HOTTech | Hot Operating Temperature Technology program |
| HWO | Habitable Worlds Observatory |
| ILEWG | International Lunar Exploration Working Group |
| IMEWG | International Mars Exploration Working Group |
| JPL | Jet Propulsion Laboratory |
| JWST | James Webb Space Telescope |
| LPSC | Lunar and Planetary Science Conference |
| MAVEN | Mars Atmosphere and Volatile Evolution Mission |
| MEPAG | Mars Exploration Program Analysis Group |
| MExAG | Mercury Exploration Assessment Group |
| NF | New Frontiers |
| OPAG | Outer Planets Assessment Group |
| OSIRIS-REx | Origins, Spectral Interpretation, Resource Identification, and Security – Regolith Explorer |
| *OWL* | *Origins, Worlds, and Life* Decadal Survey |
| PESTO | Planetary Exploration Science Technology Office |
| MatISSE | Maturation of Instruments for Solar System Exploration |
| PICASSO | Planetary Instrument Concepts for the Advancement of Solar System Observations |
| R&A | Research and Analysis |
| MEP | Mars Exploration Program |
| SAR | Synthetic Aperture Radar |
| SAW | Study Analysis Workgroup |
| SBIR | Small Business Innovation Research program |
| SIMPLEx | Small Innovative Missions for Planetary Exploration |
| TPS | Thermal Protection System |

| | | |
|---|---|---|
| **TRL** | | Technology Readiness Level |
| **UOP** | | Uranus Orbiter and Probe |
| **VEP** | | Venus Exploration Program |
| **VERITAS** | | Venus Emissivity, Radio Science, InSAR, Topography, and Spectroscopy |
| **VeSCoor** | | Venus Science Coordination group |
| **VEXAG** | | Venus Exploration Analysis Group |
| **VFM** | | Venus Flagship Mission |
| **VISE** | | Venus In Situ Explorer |
| **VISTA** | | Venus In Situ Sample Transfer and Analysis |
| **WISPR** | | Wide-Field Imager for Parker Solar Probe |

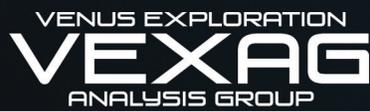

© 2024 Venus Exploration Analysis Group

Rev. 2024.12/01